\documentclass[twocolumn,aps,prl,groupedaddress]{revtex4-1}
\usepackage{graphicx}
\usepackage{xcolor}
\usepackage{amsmath}
\usepackage{enumitem}

\newcommand{\be}{\begin{equation}}
\newcommand{\ee}{\end{equation}}

\newcommand{\bea}{\begin{eqnarray}}
\newcommand{\eea}{\end{eqnarray}}

\renewcommand{\vec}[1]{{\bf #1}}
\usepackage{amssymb, amsmath}

\begin{document}
\title{Alternating currents and shear waves in viscous electronics}
\author{M. Semenyakin$^{1,2,3}$ and G. Falkovich$^{4,5}$}

\affiliation{$^1$ Center for Advanced Studies, Skolkovo Institute of Science and Technology, Moscow, Russia\\$^2$ Taras Shevchenko National University of Kyiv, Kyiv, Ukraine\\$^3$ Department of Mathematics, NRU HSE, Moscow, Russia \\$^4$  Weizmann Institute of Science, Rehovot, Israel\\$^5$Institute for Information Transmission Problems, Moscow, Russia}

\date{\today}

\bibliographystyle{abbrv}

\begin{abstract}
Strong interaction among charge carriers can make them move like viscous fluid. Here we  explore alternating current (AC) effects in viscous electronics.  In the Ohmic case, incompressible current distribution in a sample adjusts fast to a time-dependent voltage on the electrodes, while in the viscous case, momentum diffusion makes for retardation and for the possibility of propagating slow shear waves.
We focus on specific geometries that showcase interesting aspects of such waves: current parallel to a one-dimensional defect  and current applied across a long strip.
We find that the phase velocity of the  wave propagating along the strip respectively increases/decreases with the frequency for no-slip/no-stress boundary conditions.  This is so because when the frequency or strip width goes to zero (alternatively, viscosity go to infinity), the wavelength of the current pattern tends to infinity in the no-stress case and to a finite value in a general case. We also show that for DC current across a strip with no-stress boundary, there only one pair of vortices, while there is an infinite vortex chain for all other types of boundary conditions. 
\end{abstract}
\maketitle

Existence of strongly interacting carriers in high-mobility materials opens fascinating possibilities for "viscous electronics" where current flows like a viscous fluid rather than according to Ohm's law \cite{bandurin2015,crossno2016,moll2016,Torre2015,LF}. Here we describe a new phenomenon that could be observable in such materials - propagating shear waves. We show that, apart from intrinsic interest, observing such waves gives an independent way to measure the viscosity of the electronic fluid and establish what are the real boundary conditions satisfied by electronic flows.

Propagation of weak low-frequency currents in strongly interacting systems is described by classical viscous hydrodynamics. Viscous hydrodynamics has been mostly focused on the flows past the bodies. Viscous electronics makes it necessary to consider flows produced by sources and sinks. Studies of DC currents were started recently in \cite{LF,FL,Torre2015} and brought several interesting effects (current flowing against electric field, super-ballistic conductance, electric field expulsion from a flow, etc), some of which were observed experimentally \cite{bandurin2015,SB2017}.

Here we present a study of alternating current (AC)  either flowing across the strip or past the obstacles like strongly disordered zones in the bulk. 
In the Ohmic case, time-dependent voltage on the electrodes makes the current and potential distribution instantaneously adjust to it, as far as the flow is incompressible. In the viscous case, momentum propagates by diffusion, which leads to retardation and the possibility of running waves. We consider only charge neutral viscous modes, interaction of electromagnetic waves with viscous electron flows was considered in \cite{EM}.


For weak currents we can neglect non-linearity in the Navier-Stokes equations and by incorporating Ohmic resistance get the following equation:
\begin{equation}
mn(\partial_t+\gamma_p(r))v_i-\eta \nabla^2 v_i = -ne\partial_i \varphi.
\end{equation}
For AC case all quantities depend on time as $e^{-i\Omega t}$. For such dependence equation gets form:
\begin{equation}
(-i\Omega+\gamma_p(r)) v_i- \nu\nabla^2 v_i = -em \partial_i\varphi\,,\label{NS1}
\end{equation}
where $\nu=\eta/mn$ is the kinematic viscosity.

We start by describing the simplest setting for generating a shear wave. During the process of placing a graphene sheet on an insulating substrate many impurities are accumulating between  them. Due to Van-der-Waals forces, the impurities tend to concentrate in the localized regions, "bubbles" and "folds", where  resistance is high. Running AC current through the sample with such regions will generate shear viscous waves transversal to the current. If impurities concentrate in a long fold, we suggest running AC current parallel to its boundary $\vec{v}_0(x,y) = v_0 e^{-i\Omega t} \vec{e}_x$. We assume that $\gamma_p\to\infty$ inside the fold and zero outside. Then the current must turn to zero at the boundary of the current-carrying region, which thus corresponds to the no-slip boundary condition. The solution of (\ref{NS1}) then has a simple form
\begin{equation}
v_x(y,t)=v_0\,\mathrm{Re}\left(\exp[-i\Omega t](1-\exp[-|y|\sqrt{-i\Omega/\nu}])\right),
\end{equation}
which describes a wave propagating with the speed $\sqrt{2\nu\Omega}$ while oscillating and exponentially decreasing in space with the same wavenumber $\kappa=\sqrt{\Omega/2\nu} $. Therefore, registering such a wave gives one an ability to directly measure the viscosity of the electronic fluid. The above consideration is valid at sufficiently low frequencies 
such that the speed of the viscous wave is much smaller then the speed of sound-plasmon mode: $\sqrt{2\nu\Omega}\ll v_F/\sqrt{2}$. On the other hand, the wavelength must be less than the sample size, which is realistically not much larger than $Nl_{ee}$ with $N\simeq 5$.  One can estimate $\nu\simeq v_Fl_{ee}$ where $l_{ee}$ is the mean free path for momentum-conserving electron-electron collisions. That allows one to recast the applicability condition as $ N^{-2}<\Omega l_{ee}/ v_F<1$. At $\Omega\simeq 10GHz$ the respective wavelength is several microns for graphene with $v_F\simeq 10^6\,m/sec$ and $\nu\simeq 10^3cm^2/sec$ \cite{bandurin2015}. Due to small sizes of samples, the retardation effects for EM waves related to the finite light speed can be neglected up to THz frequencies, for which EM wave length is about $100\;\mu m$. 

In order of increasing complexity, we consider now the current injected into a half plane. The potential in the half-plane with a no-stress boundary can be computed exactly:
$$
\phi=\dfrac{I_0\nu}{\pi m e}\mathrm{Re}\, e^{-i\Omega t}\left(\frac{\gamma_p-i\Omega}{\nu}\log (r\lambda_{\mathrm{IR}})-\frac{y^2-x^2}{\left(x^2+y^2\right)^2}\right)
$$
Note that finite frequency is equivalent to a finite imaginary resistivity, for the large $r$-s - we have logarithmic behavior  at infinity as in the Ohmic case. And like in the usual electrical networks, "impedance" $z=\dfrac{\gamma_p-i\Omega}{\nu}$ defines the phase shift between $I$ and $\phi$. But in half-plane there are no real running waves of the potential - only zero-potential line which is oscillating between $0$ and $\infty$ once each half-period. In the no-slip case we have the same asymptotic behaviour. However, running waves could be clearly seen on the vorticity map. For example, vorticity for the no-stress case is given by:
$$
\omega=-\dfrac{I}{\pi}\mathrm{Re}\, e^{-i\Omega t}\int_{0}^{+\infty}e^{-q y} k \sin(kx)dk
$$
where $q^2=k^2 + \varkappa$, $\varkappa=(\gamma_p-i \Omega)/\nu =\rho e^{i\theta}$ so $\rho$ describes overall intensity of resistance, and $\theta=-\arctan{\Omega/\gamma_p}$ - relative contributions of reactance and resistance. As $\Omega>0,\; \gamma_p>0$, thus $ 0>\theta>-\pi/2$. Properties of the running wave can be extracted by considering the asymptotic $y\to+\infty$ in the vicinity of $x=0$, where the integral oscillates and exponentially decreases with $y$. Vorticity in this limit is given by:
$$\omega= -I x \sqrt{\dfrac{\rho^{3/2}}{2\pi y^{3}}} \cos\left(-y\sqrt{\rho}\sin{\frac{\theta}{2}}+\frac{3}{2}\theta-\Omega t\right)e^{-y\sqrt{\rho}\cos{\theta/2}}$$
The propagation speed of zero-vorticity lines and the amplitude decay rate are the same in both no-slip and no-stress cases and respectively given by
\begin{equation}
\label{scales}
v=\dfrac{\Omega}{\sqrt{\rho}|\sin{\theta/2}|}, \;\;\; \gamma = \sqrt{\rho}\cos{\frac{\theta}{2}}.
\end{equation}
The main difference between no-slip and no-stress cases is in the behavior near the boundary. In the no-stress case zero vorticity lines are approaching edge in the transverse direction, while in the no-slip case they are oriented along the edge. Running waves and behaviour near the boundary can be see in fig. 1 and fig. 2 in Supplementary Materials). Similar difference in the behaviour near the boundary could be also observed in the strip geometry.

Let us now describe in detail how AC current across the strip generates a shear wave running along the strip. It is instructive to comment on the DC case first. In this case, at the distance from the electrodes comparable to the strip width $w$, the pair of separatrices appears, dividing the inside streamlines connecting electrodes  and closed lines outside, that belong to vortices \cite{FL}. The pattern of the vortical flow outside depends crucially on the boundary conditions \cite{S}. If the boundary is stress-free then the streamlines close to the separatrices are able to go arbitrary far before turning back. If, however, boundary stress is non-zero (as, for instance, at a no-slip boundary) then the streamlines turn back at a finite distance and a chain of vortices appears (an infinite chain in an infinite strip). The properties of waves in the AC case are then also strongly dependent on the boundary conditions, as shown below.

Experimentally, it is most feasible to change the frequency $\Omega$. Whether the frequency is large or small is determined by comparing the period with the viscous time of momentum diffusion across the strip, $\tau=w^2/\nu$. Therefore, the dimensionless parameter is $\Omega\tau=\Omega w^2/\nu$. We can also denote $D_v=\sqrt{\nu/\Omega}$, which is the characteristic vortex's length scale, as it can be seen e.g. from the formula (\ref{scales}). Since $\Omega\tau = (w/D_v)^2$ then low frequency (DC limit) corresponds to a narrow strip. When the frequency $\Omega\to 0$, we find very different phase velocities for different boundary conditions: no-slip boundary corresponds to the wave velocity going to zero while no-stress to a finite value.

Assuming translational symmetry along $x$ and a uniform Ohmic resistance we write for the Fourier harmonics of the stream function defined by $\vec{v}(x,y)=\nabla \times \vec{e}_z\psi(x,y)$:
\begin{equation}
(\partial_{yy}-q^2)(\partial_{yy}-k^2) \psi=0
\end{equation}
Considering dynamics of the fluid constrained by the two edges at $-w/2$ and $w/2$, we have to put also boundary conditions. The velocity component normal to the boundary is zero everywhere, except the source and the sink: $v_y(x,\pm w/2)=I_0 \delta(x)$. For the tangential component we generally impose mixed conditions $v_x=l \partial_y v_x$ at $y=-w/2$ and $v_x=-l \partial_y v_x$ at $y=w/2$, which transforms into no-slip in the limit $l\to 0$ and into no-stress in the limit $l \to \infty$. Dependence of the results on $l$ could be analytically evaluated in the DC case (see Supplementary Materials): for a finite nonzero $l$ the features are qualitatively similar to the no-slip limit. Influence of a finite Ohmic resistance is similar to that in the half-plane case, so from now on we neglect Ohmic resistance.\\

We start analysis of vortex dynamics in the strip from the vorticity distribution in the no-stress case:
\begin{equation}
\label{stripNostrOmega}
\omega (x,y)=-\dfrac{I}{\pi}\mathrm{Re}\;e^{-i\Omega t}\int\limits_{0}^{+\infty} k \sin{k x} \dfrac{ \cosh{yq}}{ \cosh{\frac{wq}{2}}}dk
\end{equation}
For a wide strip, dynamics is almost width-independent: vortices are ejected from the electrodes and move toward the mid-line of the strip, where they meet, join, and move along the strip as a single big vortex which occupies entire strip. It can be seen, that far from the source they have regular form, distinct geometrical periodicity and on average - vorticity decays exactly exponentially.

\begin{figure}[h]
\label{noresomega}
\begin{center}
\includegraphics[width=0.5\textwidth]{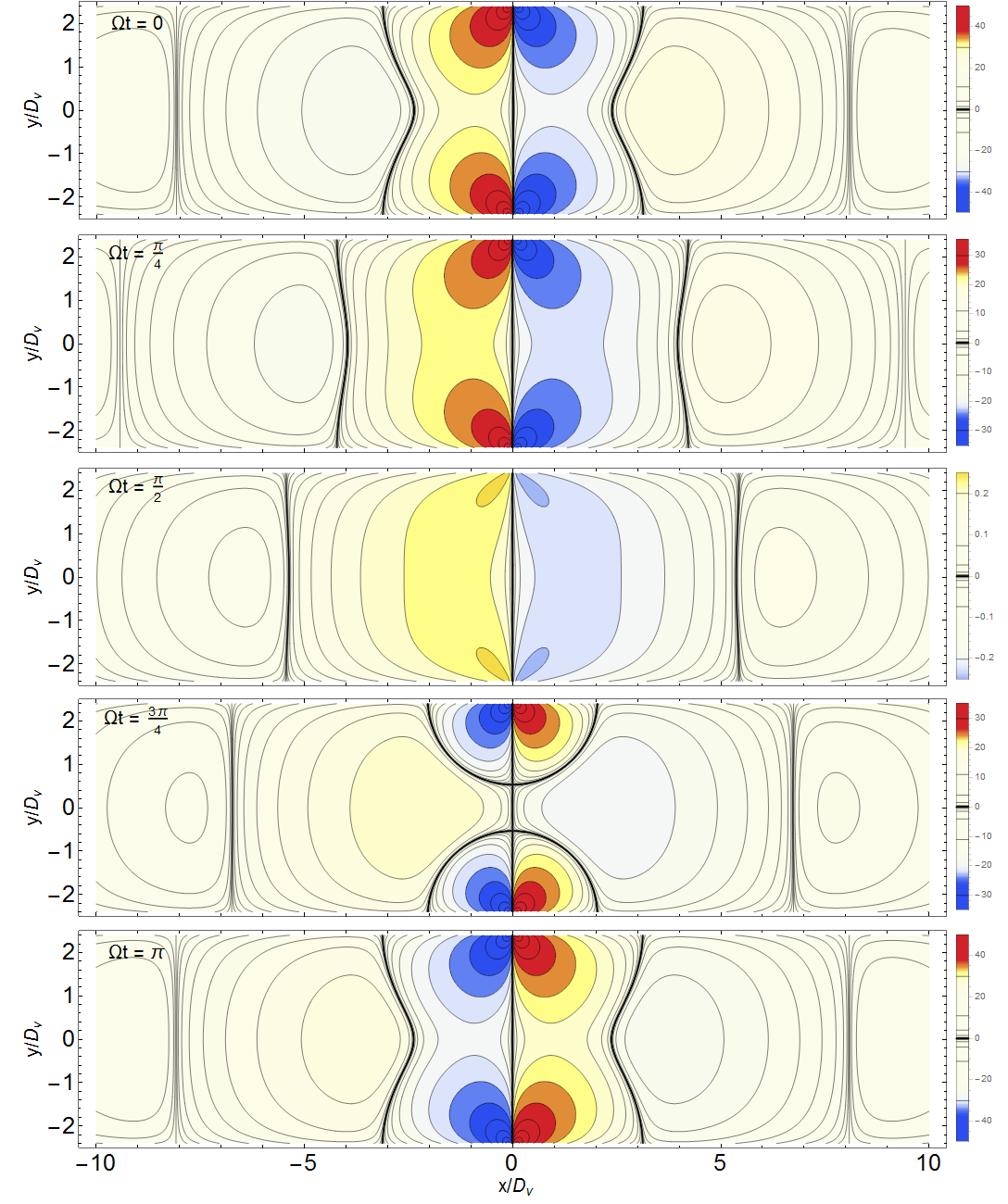}\\
\caption{Contours of constant vorticity $\omega(x,y)=const$ for the no-stress boundary conditions for $w/D_\nu=5$. Different pictures correspond to different moments of time. Places where lines condensate correspond to the isolines $\omega(x,y)=0$. Videos with the dynamic here and below can be sent by authors on demand.}
\end{center}
\end{figure}
When $w/D_\nu\to\infty$, the distance between vortices saturates to a constant, while if $w/D_\nu\to 0$, the wave length tends to infinity as $D_\nu/w$. The results of numerical computation shown in the Figure~2.a are in a good agreement with the results of the "saddle point" estimation for the integral. To put it simply, vortices cannot be squeezed into too narrow strip.
\begin{figure}[h]
\label{laonom}
\begin{center}
\includegraphics[width=0.5\textwidth]{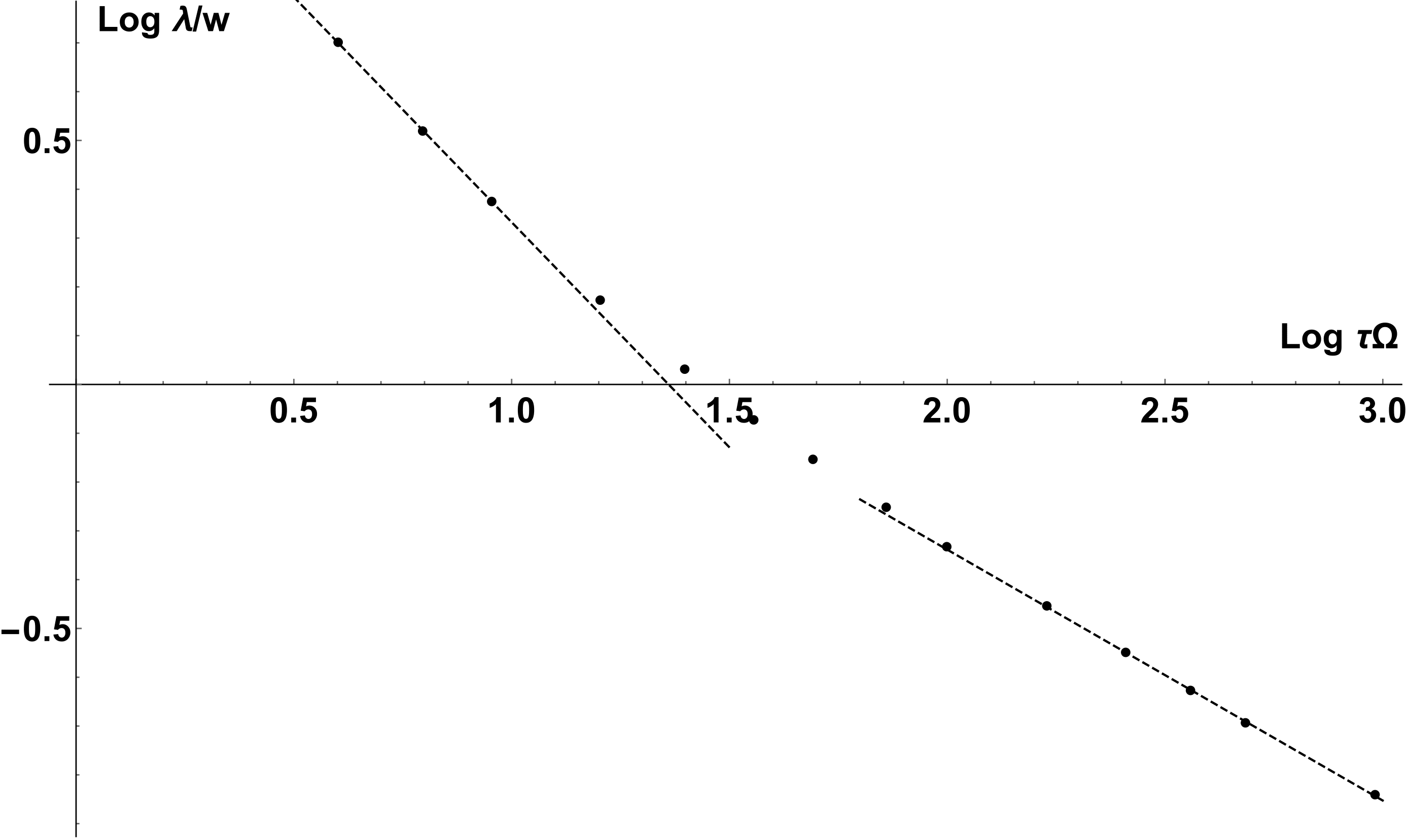}
\includegraphics[width=0.5\textwidth]{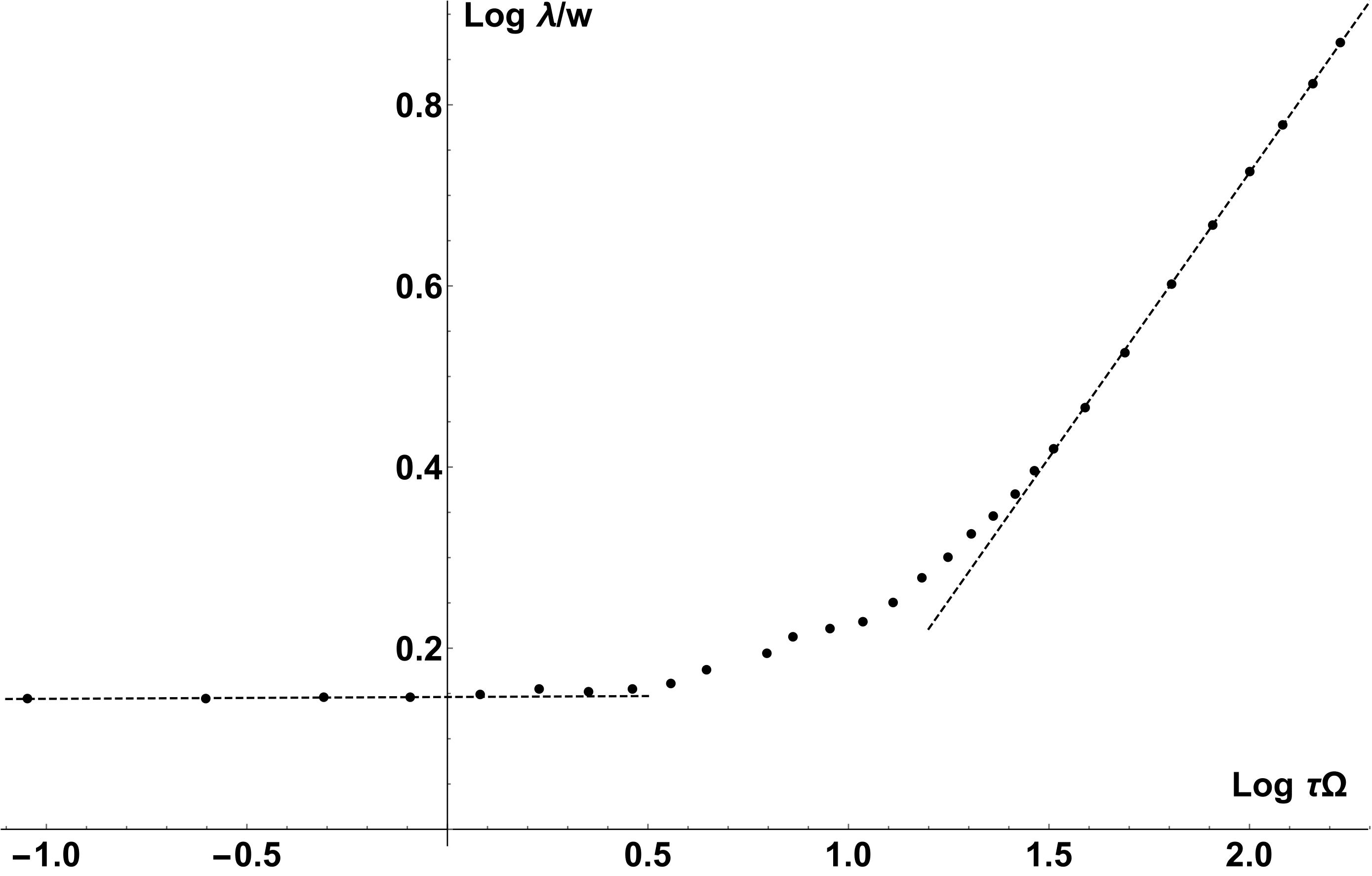}
\caption{Dependence of the distances between zeroes of function $\omega(x,0)$ (see fig. 3) far from the source in log-log scale. Lines - linear fit, points - results of numerics. Upper panel: no-stress case. Asymptotic dependence on the frequency is $\lambda/w\sim (\tau\Omega)^{-1}$, $\Omega\to 0$ and $\lambda/w\sim (\tau\Omega)^{-1/2}$, $\Omega\to \infty$  ($x\gg w,\;D_\nu$) Lower panel: no-slip case. Here $\lambda/w\sim const$, $\Omega\to 0$ and $\lambda/w\sim (\tau\Omega)^{a}$, $\Omega\to \infty$, $a\sim 0.6$. Characteristic time scale is $\tau=w^2/\nu$}
\end{center}
\end{figure}

Let us see how different is the no-slip case. Vorticity in this case is given by:
$$\omega=\dfrac{I}{\pi}\;\mathrm{Re}\; e^{-i\Omega t}\!\int\limits_0^{+\infty}\!\dfrac{dk\;\varkappa\cosh{y q}\sinh{\frac{wk}{2}}\sin{k x}}{k\cosh \frac{wq}{2} \sinh \frac{wk}{2} - q \cosh \frac{wk}{2} \sinh \frac{wq}{2}}$$
The major striking difference is that zero-vorticity lines at mid-strip move towards the source as seen in fig. 3. The reason is that the wave of vorticity is emitted from the source not as a round vortices, as in the no-stress case, but rather as elongated ellipses oriented along the edge.
\begin{figure}[h]
\begin{center}
\includegraphics[width=0.5\textwidth]{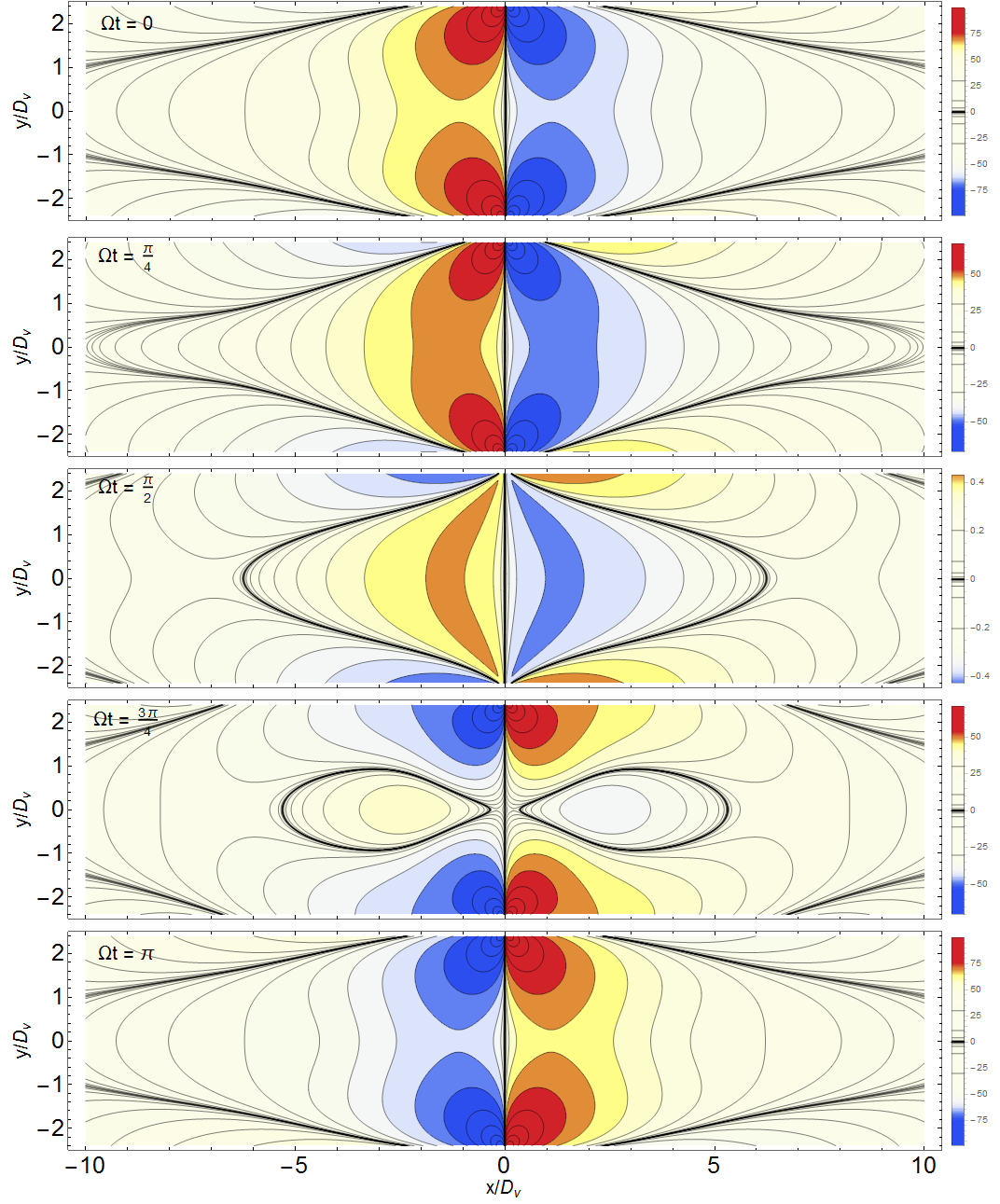}
\caption{Contours of the constant vorticity $\omega(x,y)=const$ for the no-slip boundary conditions for $w/D_\nu=5$.}
\end{center}
\end{figure}

The movement of the vortex line in mid-strip is the result of the meeting of 'waves', coming from the source and the sink. Frequency dependence of the distance between zero-vorticity points at $y=0$, which in fact is the vortex size, is shown on the picture 2.b. This horizontal distance is different from the vertical distance between "layers" in the half-plane case, and thus doesn't tend to some constant in the limit $w\to \infty$ (in distinction from the distance between vortices in the no-stress case). DC limit  $\Omega\to 0$ corresponds to $w/D_\nu\to 0$. As follows from the consideration of the DC case, there must still exist vortices of finite length in this limit, as long as
\begin{equation}
\lambda/D_\nu\sim w/D_\nu \;\;\Rightarrow \lambda \sim w,
\end{equation}
that is the vortex size shouldn't depend on $\Omega$. 
Another interesting novelty in comparison with the no-stress case is that for the narrow strip vortexes are moving by jumps, not smoothly. The less is $w$, the shorter is the duration of jump - major part of the period vortices are spending as a standing wave, and only when the amplitude is very little, $I(t) \to 0$, they are moving. In the limit $w\to \infty$, vortexes are moving smoothly. This phenomenon is due to the asymmetry between the real and imaginary parts of the function.

Formula for the voltage in the no-stress case has the form:
$$
V_{l\to \infty}(x,y)=\phi(x,y)-\phi(+\infty,y)=
$$
\begin{equation}
=-\dfrac{m\nu I_0}{e\pi}\;\mathrm{Re}\;e^{-i\Omega t}\int\limits_{0}^{+\infty} dk\dfrac{q^2 \sinh ky}{k\cosh \frac{kw}{2}}\cos{k x}
\end{equation}
and, surprisingly, there are no vortices at all, as it can be seen, e.g. on the fig.4. This happening because of miraculous cancellation of the terms containing $\cosh{qw/2}$ in the numerator and in the denominator. However, this cancellation is absent for finite $l$ length, and thus for general $l$ we shall see vortices as in the no-slip case.
\begin{figure}[h]
\begin{center}
\label{nostrvolt}
\includegraphics[width=0.45\textwidth]{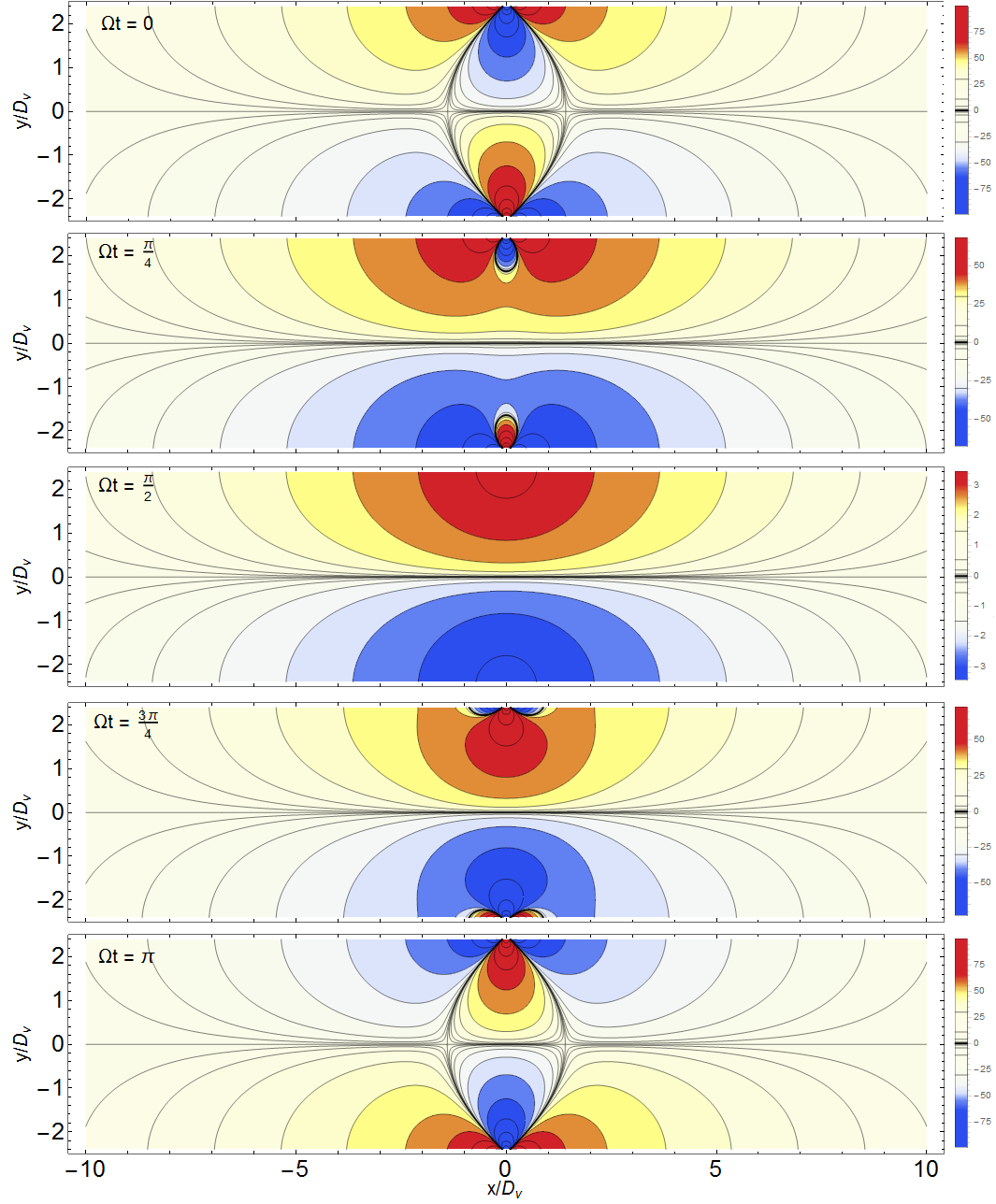}
\caption{Distribution of the voltage for the no-stress case}
\end{center}
\end{figure}
Potential for the no-slip case has the form:
$$V_{l\to 0}(x,y)=\phi(x,y)-\phi(+\infty,y)=-\dfrac{m\nu I_0}{e\pi}\;\mathrm{Re}\;\varkappa e^{-i\Omega t}\times$$
\begin{equation}
\times\int\limits_{0}^{+\infty} \dfrac{qdk}{k}\dfrac{\sinh ky \sinh \frac{wq}{2} \cos{k x}}{k\cosh \frac{wq}{2} \sinh \frac{wk}{2} - q \cosh \frac{wk}{2} \sinh \frac{wq}{2}}
\end{equation}
General behaviour is qualitatively similar to that of the vorticity, including freezing and inverted phase speed.
\begin{figure}[h]
\begin{center}
\includegraphics[width=0.45\textwidth]{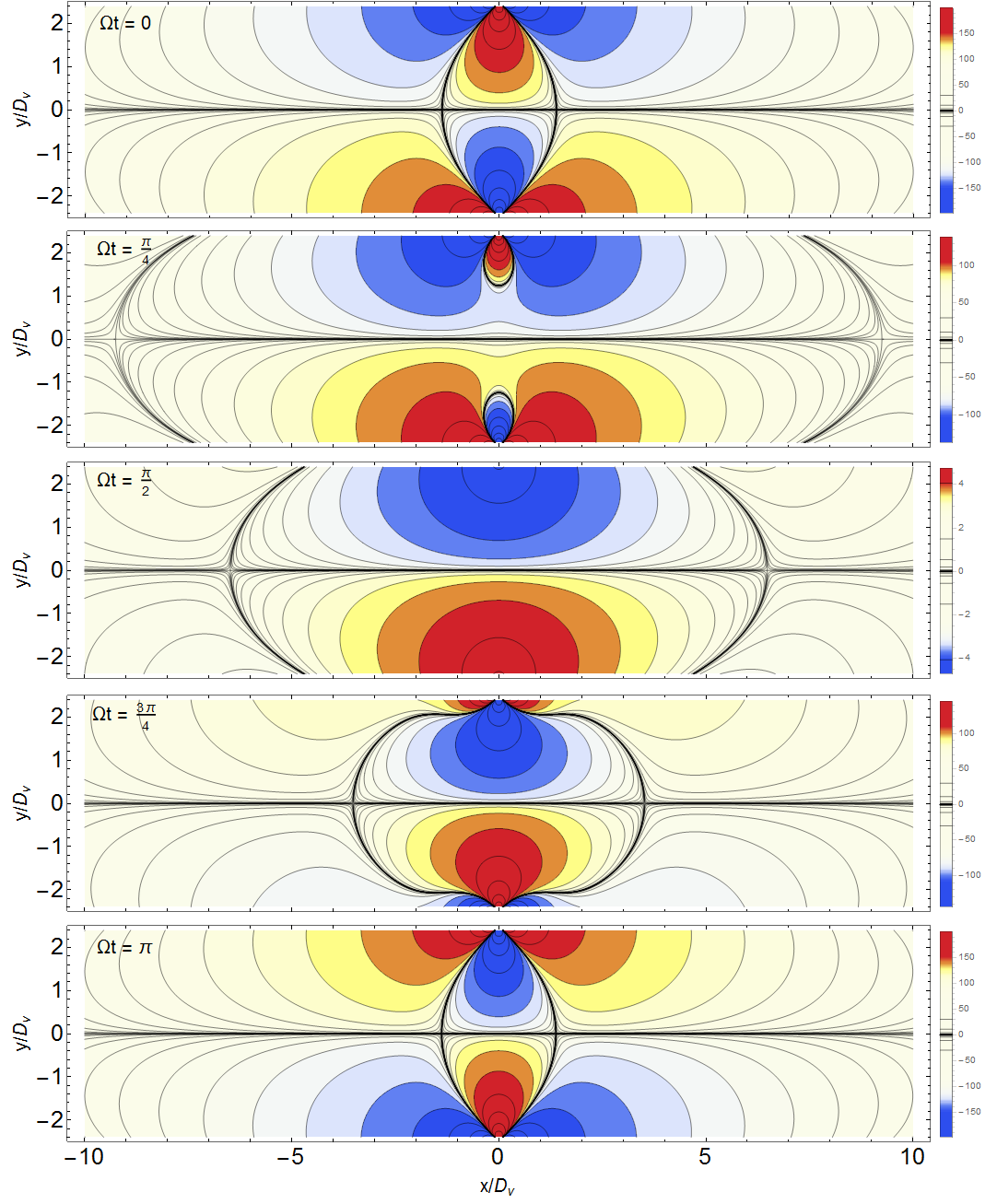}
\caption{Distribution of the potential in the no-slip case.}
\end{center}
\end{figure}

To conclude, vorticity and potential waves propagating along the strip
are qualitatively different for no-stress and no-slip boundary conditions - waves could be observed on the vorticity map in the both cases, and on the potential map in the no-slip case only. There is no running potential wave  for no-stress case. Moreover, phase speed of the waves is directed in the opposite directions in the different cases. Wave-length of viscous waves also depending on the frequency in a different ways for the no-stress and no-slip cases.

We are much grateful to Leonid Levitov for numerous helpful discussions, his input was indispensable for this work. We acknowledge support of the
MISTI MIT-Israel Seed Fund, Minerva Foundation, 
the  Israeli Science Foundation (grant 882) and the Russian Science Foundation (project 14-22-00259). M.S. wants to thank Oleksandr Gamayun and Pavlo Gavrylenko for permanent support during the work.


\onecolumngrid
\section*{Appendix A. AC current in a half-plane}

To find the velocity field of a Ohmic-viscous flow we need to solve the equation:
\begin{equation}
(-i\Omega+\gamma_p(r)) v_i- \nu\nabla^2 v_i = -em \partial_i\varphi\,,\label{NS1}
\end{equation}
with the uniform Ohmic resistivity $\gamma_p(r)=const$. Acting by $\nabla \times$ on both sides  and assuming translational invariance along $x$, we get the equation for the stream function:
\begin{equation}
(\partial_{yy}-q^2)(\partial_{yy}-k^2) \psi=0.
\end{equation}
Here the stream function is defined by $\vec{v}(x,y)=\nabla \times \vec{e}_z\psi(x,y)$, its Fourier image $\psi(x,y)=\int\dfrac{dk}{2\pi} e^{ikx}\psi(k,y)$, and
$$q^2=k^2 + \varkappa,\;\;\; \varkappa=(\gamma_p-i \Omega)/\nu =\rho e^{i\theta}.$$
This equation has 4 solutions $\psi(k,y)=c_1 e^{-|k| y} + c_2 e^{|k|y} + c_3 e^{q y} + c_4 e^{-qy}$. General boundary conditions $v_y(x,0)=I_0 \delta(x)$ and $v_x(x,0)=l\partial_y v_x(x,0)$ give $\psi(k,0)=I/(ik)$ and $l\partial_{yy} \psi(k,0) = \partial_{y} \psi (k,0)$. Adding condition $v(x,y)\to 0, \;\; y\to +\infty$ we completely define all the coefficients and find:
\begin{equation}
\psi(x,y) = \dfrac{I}{\pi} \int\limits_{0}^{+\infty}\dfrac{ e^{-q y} k (1 + k l) - e^{-k y} q (1 + ql)}{k (k-q) (1 +  l (k + q))}\sin(kx)dk\ .
\end{equation}
The solution  for the non-resistive case corresponds to the limit $\varkappa \to 0$. The vorticity  $\omega(x,y)=\Delta \psi(x,y)$ is plotted in Figure 6, where one can see  two vortices appearing every half-period. A line of zero vorticity separates the two vortices from the next pair. In the lower panel of Figure 6, the vorticity is shown for the case with strong ohmic resisitivity, $\theta=-\pi/6$. It can be seen that in this case vortices disappear much faster, yet there are no qualitative differences. Thus, for simplicity sake, further we will consider non-ohmic case only. In the no-slip case, zero-vorticity line is oriented along the edge of the bulk. On the contrary, the line comes in the transverse direction in the no-stress case. This difference gives qualitatively different pictures for the strip case.\\
Most of the flow properties related to vortices are encoded in the vorticity:
\begin{equation}
\omega(x,y)=-\dfrac{I}{\pi} \int\limits_{0}^{+\infty}\dfrac{ (1 + k l)(k+q)}{1 +  l (k + q)}e^{-q y} \sin(kx)dk
\end{equation}
In the no-stress case $l \to +\infty$ we have:
\begin{equation}
\label{nostres}
\omega(x,y)=-\dfrac{I}{\pi} \int\limits_{0}^{+\infty}e^{-q y} k \sin(kx)dk
\end{equation}
It is vanishing at the $x=0$. However, we can consider its behaviour at $x\sim 0$. As for $k\gg 1/y, k\gg 1/\sqrt{\rho}$ exponential suppresses other integrands, and for small $x$ we can expand $\sin(kx)\sim kx$. At the first order we have:
\begin{equation}
\omega(x,y) = -\dfrac{I x}{\pi} (\partial_{yy}^2-\varkappa) \int\limits_0^{+\infty}e^{-q y}dk
\end{equation}
Obtaining asymptotic at $y\to +\infty$, in the lowest order in $1/y$ we get:
\begin{equation}
\omega(x,y)= -I x \sqrt{\dfrac{\varkappa\sqrt{\varkappa}}{2\pi y^{3}}} e^{-y \sqrt{\varkappa}},\;\;\omega(x,y,t)= \mathrm{Re}\,\omega(x,y) e^{-i\Omega t}= -I x \sqrt{\dfrac{\rho^{3/2}}{2\pi y^{3}}} \cos\left(-y\sqrt{\rho}\sin{\frac{\theta}{2}}+\frac{3}{2}\theta-\Omega t\right)e^{-y\sqrt{\rho}\cos{\theta/2}}
\end{equation}
Thus, zero-vorticity lines correspond to:
\begin{equation}
-y\sqrt{\rho}\sin{\frac{\theta}{2}}+\frac{3}{2}\theta-\Omega t=\pi\left(k+\frac{1}{2}\right),\;\;\;k\in\mathbb{Z}
\end{equation}
Or for the non-Ohmic case:
\begin{equation}
y\sqrt{\Omega/2\nu}-\Omega t=\pi\left(k+\frac{3}{4}\right),\;\;\;k\in\mathbb{Z}
\end{equation}
\begin{center}
\begin{figure}[th]
\includegraphics[width=1\textwidth]{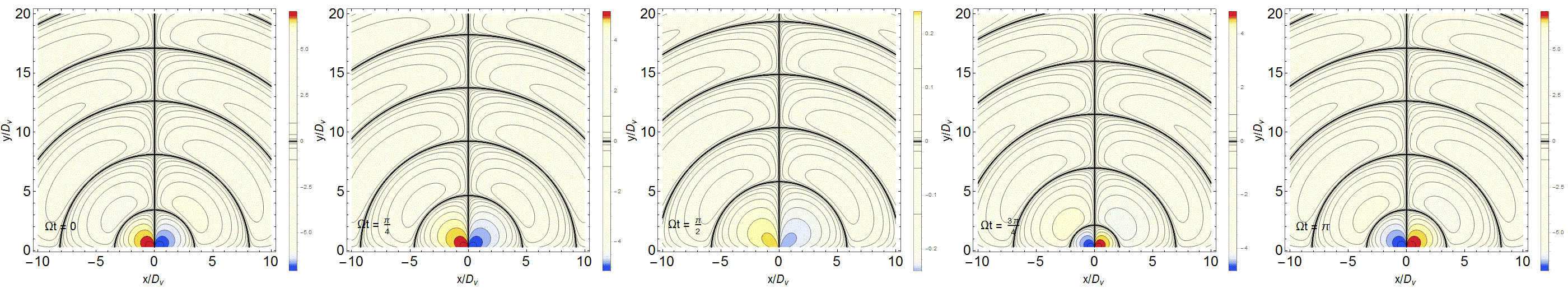}
\includegraphics[width=1\textwidth]{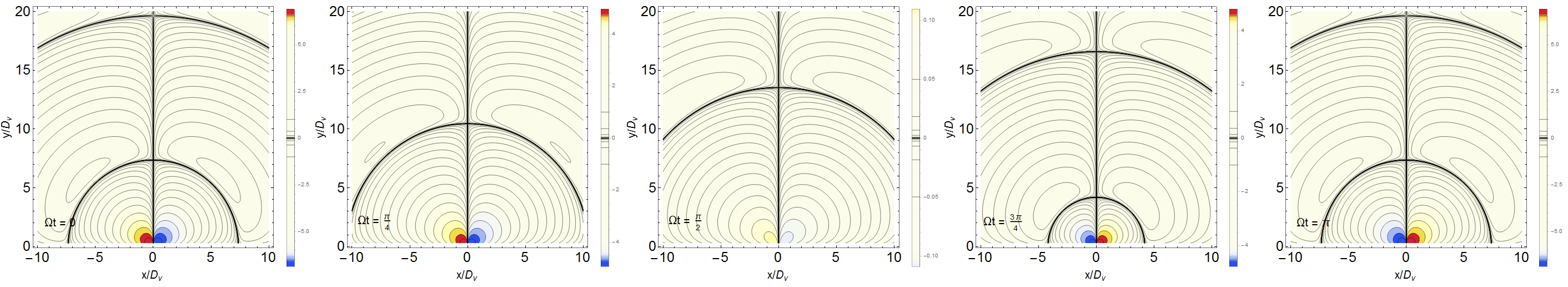}
\caption{No-stress case. Upper row: vorticity for the half-plane for the pure AC case, $\theta=-\pi/2$. Lower row: partially ohmic case: $D_\nu=1/\sqrt{\rho},\; \theta=-\pi/6$}
\end{figure}
\end{center}
In the no-slip case $l\to 0$ we have:
\begin{equation}
\label{noslip}
\omega(x,y)=-\dfrac{I}{\pi}\int\limits_0^{+\infty}e^{-q y}(k + q)\sin(k x)dk
\end{equation}
In the lowest degrees by $1/y$ and $x$ it gives:
\begin{equation}
\omega=\omega_{l \to +\infty}-\dfrac{I x\varkappa}{\pi y}e^{-y\sqrt{\varkappa}}=-I x \left( \sqrt{\dfrac{\varkappa\sqrt{\varkappa}}{2\pi y^{3}}} +\dfrac{\varkappa}{\pi y}\right)e^{-y\sqrt{\varkappa}}
\end{equation}
For large enough $y$ we get:
\begin{equation}
\omega(x,y)\sim - \dfrac{I x\varkappa}{\pi y}e^{-y\sqrt{\varkappa}},\;\;\;\omega(x,y,t)\sim - \dfrac{I x\rho}{\pi y}e^{-y\sqrt{\rho}\cos\frac{\theta}{2}}\cos\left(y\sqrt{\rho}\sin\frac{\theta}{2}+\Omega t-\theta\right)
\end{equation}
Thus, the coordinates of zero-vorticity lines are given by:
\begin{equation}
y\sqrt{\rho}\sin\frac{\theta}{2}+\Omega t-\theta=\pi\left(k+\frac{1}{2}\right)\;\; k\in \mathbb{Z}
\end{equation}
In the non-Ohmic case, it gives:
\begin{equation}
\label{halfplaneNoStressZeroLine}
y\sqrt{\Omega/2\nu}-\Omega t=\pi k\;\; k\in \mathbb{Z}
\end{equation}
The speed of zero-vorticity line and the decay rate of excitations are the same in both cases:
\begin{equation}
v=\dfrac{\Omega}{\sqrt{\rho}|\sin{\theta/2}|}, \;\;\; \gamma = \sqrt{\rho}\cos{\frac{\theta}{2}}
\end{equation}
which shows robustness of the result with respect to appearance of small ($\gamma_p<\omega$) Ohmic contribution, which only slightly changes angle $\theta$.
\begin{center}
\begin{figure}[th]
\includegraphics[width=1\textwidth]{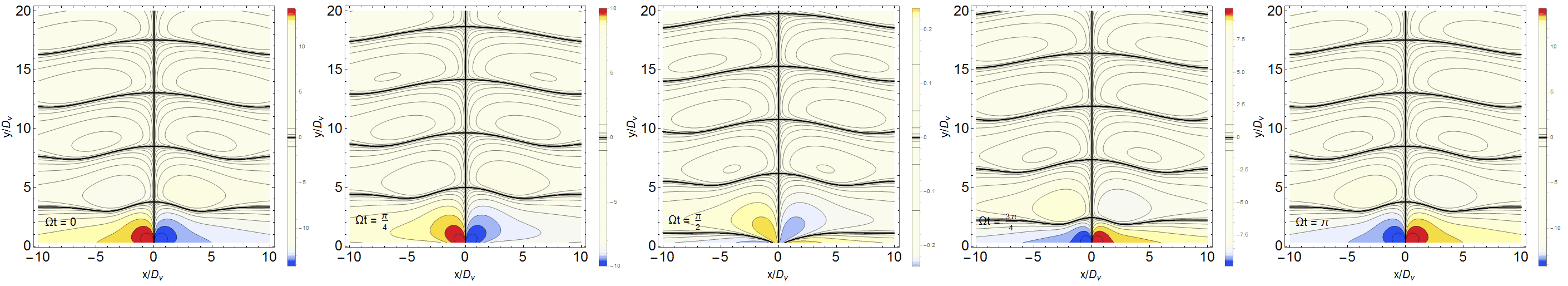}
\includegraphics[width=1\textwidth]{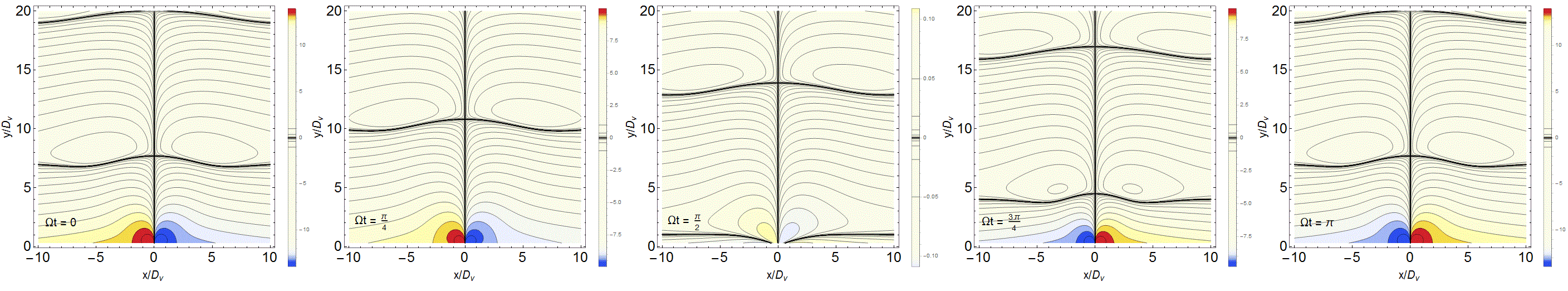}
\caption{No-slip case. Upper: vorticity for the half-plane for the pure AC case, $\theta=-\pi/2$. Lower: partially ohmic case: $D_\nu=1/\sqrt{\rho},\; \theta=-\pi/6$}
\end{figure}
\end{center}
The potential is be obtained as follows:
\begin{equation}
\partial_i\varphi = -\dfrac{\nu}{em} (\varkappa - \nabla^2) v_i
\end{equation}
which for general $l$ gives
\begin{equation}
V(x,y)=\phi(x,y)-\phi(+\infty,y)=\dfrac{\varkappa I\nu}{\pi me}\int\limits_{0}^{+\infty} \dfrac{q(1+lq)e^{-ky}}{k (k-q) (1+l(k+q))} \cos (kx) dk\ .
\end{equation}
That gives in the no-stress limit $l\to +\infty$
\begin{equation}
V_{l\to +\infty}(x,y)=-\dfrac{I\nu}{\pi me}\int\limits_{0}^{+\infty} \dfrac{q^2}{k}e^{-ky} \cos (kx) dk
\end{equation}
which has singularity at $x,y \to 0$ given by
\begin{equation}
V_{l\to +\infty}(x,y)\sim -\dfrac{I\nu}{\pi me} \dfrac{y^2-x^2}{(y^2+x^2)^2}\,.
\end{equation}
That coincides with the expression for the DC case. Asymptotic for large $x$ is given by
\begin{equation}
V_{l\to +\infty}(x,y)\sim -\dfrac{I\nu \varkappa}{\pi me}\mathrm{Re}\;\int\limits_{0}^{+\infty} \dfrac{e^{-k(y+ix)}}{k}dk\sim -\dfrac{I\nu \varkappa}{\pi me}\log (r\lambda_{IR})
\end{equation}
and has $\log$-dependence on IR cutoff (assuming that $r\ll 1/\lambda_{IR}$). This assymptotic coincide up to a complex phase with the solution for the Ohmic case. No-slip limit $l\to 0$ is as follows:
\begin{equation}
V_{l\to 0}(x,y)=-\dfrac{I\nu}{\pi me}\int\limits_{0}^{+\infty} \dfrac{q (k+q) e^{-ky}}{k} \cos (kx) dk = V_{l\to +\infty}(x,y)-\dfrac{I\nu}{\pi me}\int\limits_{0}^{+\infty} q e^{-ky} \cos (kx) dk,
\end{equation}
and has similar asymptotic behaviour. In both cases there are no running waves, as far as there is no spatially oscillating mixing between real and imaginary parts.
\section*{Appendix B. General equations for the strip}
Expression for $\psi$ in the case of strip could be obtained from the general solution
\begin{equation}
\psi(k,y)=A \cosh ky + B \cosh qy + C \sinh ky + D \sinh qy
\end{equation}
of the equation
\begin{equation}
(\partial_{yy}-k^2)(\partial_{yy}-q^2)\psi=0
\end{equation}
with the boundary conditions
\begin{equation}
\psi(k,-w/2)=\dfrac{I}{ik},\; l\partial_{yy}\psi(k,-w/2)=\partial_{y}\psi(k,-w/2)
\end{equation}
\begin{equation}
\psi(k,w/2)=\dfrac{I}{ik},\; l\partial_{yy}\psi(k,w/2)=-\partial_{y}\psi(k,w/2)
\end{equation}
(for definitions of $q$ and $\psi(k,y)$ see previous Section). General solution looks:
\begin{equation}
\psi(x,y)=-\dfrac{I}{\pi}\int\limits_{0}^{\infty} \dfrac{dk}{k}\dfrac{q \cosh ky \left(l q \cosh \frac{qw}{2} + \sinh \frac{qw}{2}\right)-k \cosh qy \left(k l \cosh \frac{kw}{2} + \sinh \frac{kw}{2}\right) }{k \cosh \frac{qw}{2} \sinh \frac{kw}{2} - q \cosh \frac{kw}{2}  \sinh \frac{qw}{2}-\varkappa l \cosh \frac{kw}{2} \cosh \frac{qw}{2}}\sin{k x}
\end{equation}
where $y$ - coordinate from the mid of the strip, and for vorticity:
\begin{equation}
\omega(x,y)=\dfrac{I}{\pi}\int dk\dfrac{\varkappa \cosh qy \left(k l \cosh \frac{kw}{2} + \sinh \frac{kw}{2}\right)}{k \cosh \frac{qw}{2} \sinh \frac{kw}{2} - q \cosh \frac{kw}{2}  \sinh \frac{qw}{2}-\varkappa l \cosh \frac{kw}{2} \cosh \frac{qw}{2}}\sin kx.
\end{equation}\\
General expression for the potential
\begin{equation}
V(x,y)=\phi(x,y)-\phi(\infty,y)=-\dfrac{I m\nu}{e\pi}\int\limits_{0}^{\infty} \dfrac{dk}{k}\dfrac{\varkappa q \sinh ky \left(l q \cosh \frac{qw}{2} + \sinh \frac{qw}{2}\right)}{k \cosh \frac{qw}{2} \sinh \frac{kw}{2} - q \cosh \frac{kw}{2}  \sinh \frac{qw}{2}-\varkappa l \cosh \frac{kw}{2} \cosh \frac{qw}{2}}\cos{k x}
\end{equation}
is computed by integration of Stokes equation in $x$ and differentiation on $y$.
\section*{Appendix C. Wavelength computations}
Dependence $\lambda/D_\nu(w/D_\nu)$ can be found analytically in the various independent ways. First of all, we can apply saddle-point approximation, as far as we have large parameter $x\to +\infty$. If we find saddle-point value of the wave-number $k_0=k_0(w)$, we will be able to obtain distance between vortices as $\lambda_0 = \pi/\mathrm{Re}\,k_0$. For integral
\begin{equation}
\omega (x,0)=-\dfrac{I}{2\pi i}\int\limits_{-\infty}^{+\infty} k e^{ikx-\ln \cosh{wq/2}}dk
\end{equation}
saddle-point equation can be written as:
\begin{equation}
\dfrac{kw}{qw}\tanh{qw/2}=i \dfrac{x}{w}\;\;\; (\mathrm{where}\;\; q=\sqrt{k^2-i/D_\nu^2})
\end{equation}
If we are interested in the behavior of the function for the large values of $x$, we should make l.h.s. of equation large. There are two ways to do so. If $w/D_\nu\gg 1$, then we need $q\ll k$, and thus $k\to (1+i)/\sqrt{2} D_\nu$, $\lambda_0=\pi\sqrt{2}D_\nu$. In the opposite limit $w/D_\nu\ll 1$ we can make $\tanh$ large by choosing $q\sim k \sim i\pi/2w+O(w/D_\nu^2)$ - main contribution is purely imaginary. Sub-leading order:
\begin{equation}
k_0 \sim i\pi/2w + \sqrt{-i}/ x + w/D_\nu^2\pi+O(w^3),
\end{equation}
in the limit $x\to +\infty$ gives $\lambda_0 = \pi^2 D_\nu^2/w$. Both results agree with the results of the numerics presented in the main text.

For the no-slip case we can again try to find $\lambda$ by finding the pole closest to the real axis. Equation for the pole has the form:
\begin{equation}
q \sinh \dfrac{qw}{2} \cosh \dfrac{kw}{2} -k\sinh \dfrac{kw}{2} \cosh \dfrac{qw}{2}=0
\end{equation}
It can be symmetrized by using replacements:
\begin{equation}
k^2=i/2(\sigma+1),\;\; \gamma=w/2\sqrt{i/2}
\end{equation}
and gets form
\begin{equation}
\sqrt{\sigma+1} \tanh \gamma \sqrt{\sigma+1}=\sqrt{\sigma-1}\tanh \gamma\sqrt{\sigma-1}
\end{equation}
In the limit $\gamma\to 0$  (or equally $w\to 0$) it can be solved by direct expansion in $\sigma$. In the first order $\sigma = 3/(2\gamma^2)$, which gives $k_0=\sqrt{6/w^2+i/2D_\nu^2},\;\;\lambda_0=\pi w/\sqrt{6}$. Opposite limit $w\to +\infty$ is treated numerically in the main text.\\

\section*{Appendix D. Comment on DC case in the strip}
The DC case in the strip was considered in the \cite{LF}, where only one pair of vortices was shown both for no-slip and and no-stress case. Here we show that there could be multiple pairs of vortices. Consider, for instance, vorticity for the DC case:
\begin{equation}
\omega(x,y)=-\dfrac{4 I_0}{\pi}\int\limits_0^{+\infty}\dfrac{k\cosh ky \left(k l \cosh \left(\frac{k w}{2}\right)+\sinh \left(\frac{k w}{2}\right)\right)}{2 kl (1+\cosh kw)+kw+\sinh kw}\sin kx\;dk
\end{equation}
At $x\gg w$ one can use the sadle-point approximatin with the following saddle point condition:
\begin{equation}
2kl (1 + \cosh{k w})+kw + \sinh{k w}=0\,.
\end{equation}
It gives in the two limits:
\begin{itemize}
\item $l \to 0$: $kw+\sinh kw=0$ - lots of solutions, with non-zero imaginary part
(minimal - with $\mathrm{Re}\; k_0\sim 2.25$).\\
\item $l \to \infty$: $k_0\sim i\pi(2k+1)+\dfrac{1}{\sqrt{lw}}$, i.e. $\lambda=\dfrac{\pi}{\mathrm{Re}\; k_0}\sim\pi \sqrt{lw}$
\end{itemize}
Results of numerical solving of this equation is given in the figure below, which coincides well with the analytic asymptotic.
\begin{center}
\begin{figure}[th]
\includegraphics[width=0.7\textwidth]{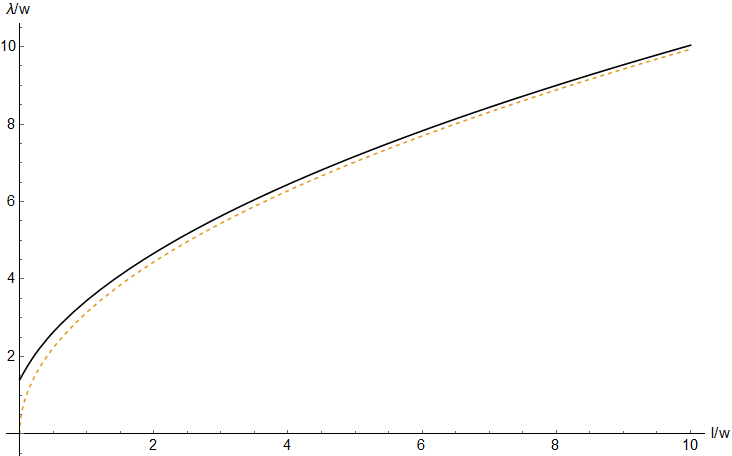}
\caption{Dependence of the wavelength on the slippage parameter. In the limit $l\to +\infty$ the wavelength tends to infinity, while for all the other values it remains finite. Dashed line - analytical asymptotic for $l\to +\infty$.}
\end{figure}
\end{center}

\end{document}